\documentclass[namedreferences]{solarphysics}

\usepackage[hyperref,optionalrh]{spr-sola-addons} 
\usepackage{graphicx}        
\usepackage{color}           
\usepackage{breakurl}        

\chardef\us=`\_

\def\referee#1{{#1}}

\begin{document}

\begin{article}
\begin{opening}

\title{Publication statistics on Sun and heliosphere}

\author{Carolus J.\ \surname{Schrijver}$^1$}

\runningtitle{Publication statistics on Sun and heliosphere}
\runningauthor{C.J.\ Schrijver}

\institute{$^1$ Lockheed Martin Advanced Technology Center,
               3251 Hanover St., Palo Alto, U.S.A., 
                     email: \url{schrijver@lmsal.com}}


\begin{abstract}
  The professional literature provides one means to review
  the evolution and geographic distribution of the scientific
  communities engaged in solar and heliospheric physics.  With help of
  the Astrophysics Data System (NASA/ADS), I trace the growth of the
  research community over the past century from a few dozen
  researchers early in the 20-th Century to over 4,000 names with over
  refereed 2,000 publications in recent years, with 90\%\ originating from 20
  countries, being published in 90 distinct journals. Overall, the
  lead authors of these publications have their affiliations for 45\%\
  in Europe, 29\%\ in the Americas, 24\%\ in Australasia, and 2\%\ in
  Africa and Arab countries. Publications most frequently appear (in
  decreasing order) in the Astrophysical Journal, the Journal of
  Geophysical Research (Space Physics), Solar Physics, Astronomy and
  Astrophysics, and Advances in Space Research (adding up to 59\%\ of
  all publications in 2015).
\end{abstract}

\end{opening}
\ifx \doiurl    \undefined \def \doiurl#1{\href{http://dx.doi.org/#1}{\textsf{DOI}}}\fi
\ifx \adsurl    \undefined \def \adsurl#1{\href{http://adsabs.harvard.edu/abs/#1}{\textsf{ADS}}}\fi
\ifx \arxivurl  \undefined \def \arxivurl#1{\href{http://arxiv.org/abs/#1}{\textsf{arXiv}}}\fi

\setcounter{footnote}{1}
\section{Trends in the research community and its publications}
\cite{comm10final} reviewed the publication statistics on research
related to Sun and heliosphere starting in 1911 and ending in 2014.
They did so on the occasion of the reorganization of the Commission
structure of the International Astronomical Union which formally ended
its Commission 10 on ``Solar Activity''.  \cite{comm10final} analyzed
the activity of the research community using the tools provided by the
Astrophysics Data System (ADS\footnote{URL:
  http://adsabs.harvard.edu/abstract$\_$service.html}), which enables
searches over all the major trade publications in astrophysics in
general. They reviewed the number of refereed publications per year
going back over a century, and quantified the population of active
researchers and their publication productivity.  In the remainder of
this Section only, I use their results and largely their wording and
conclusions, but updated the figures and other results to include
2015 and added comments on the trend changes in the mid-1970s.

\begin{figure}[t]
\centering
\includegraphics[width=0.9\textwidth]{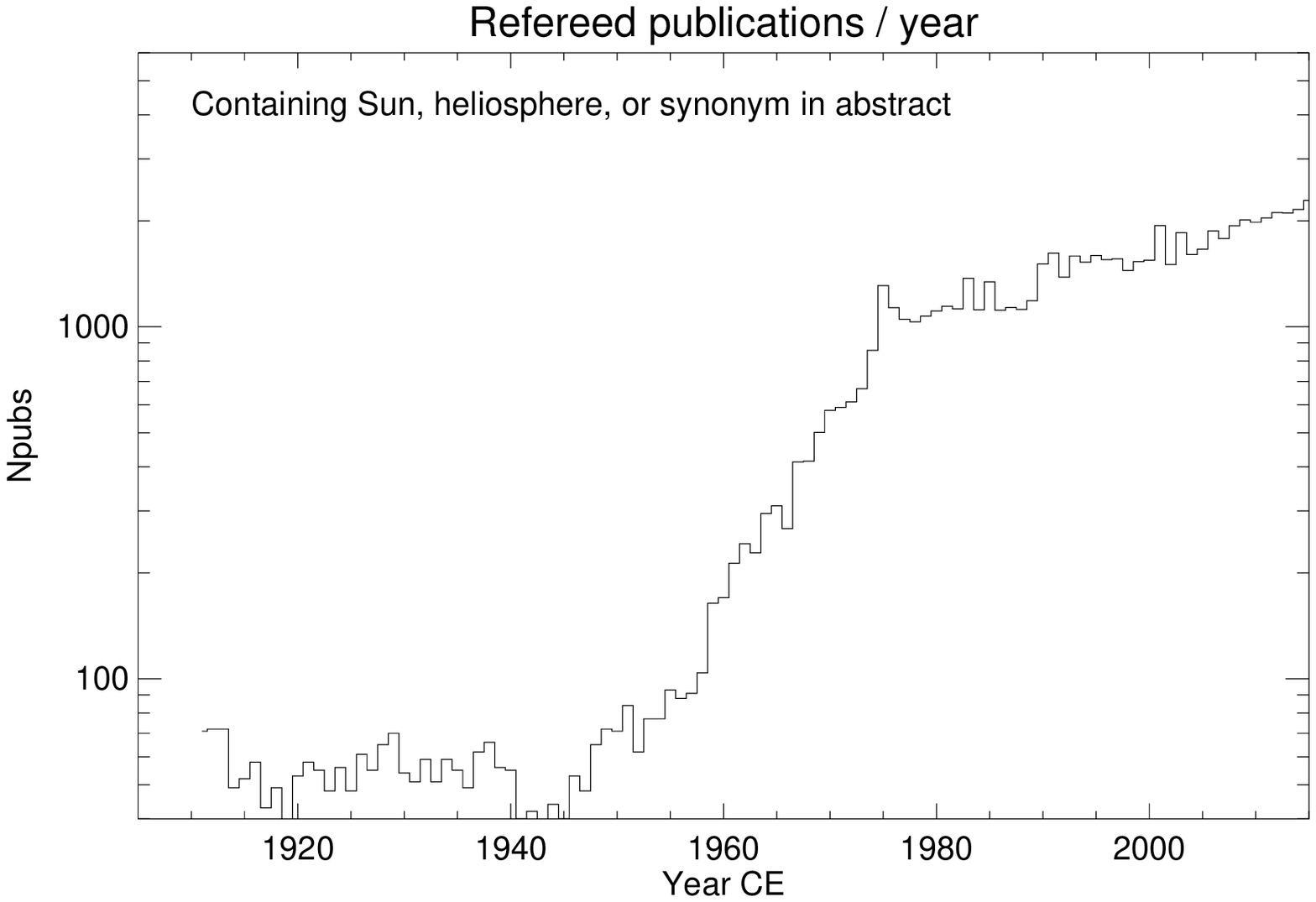}
\includegraphics[width=0.9\textwidth]{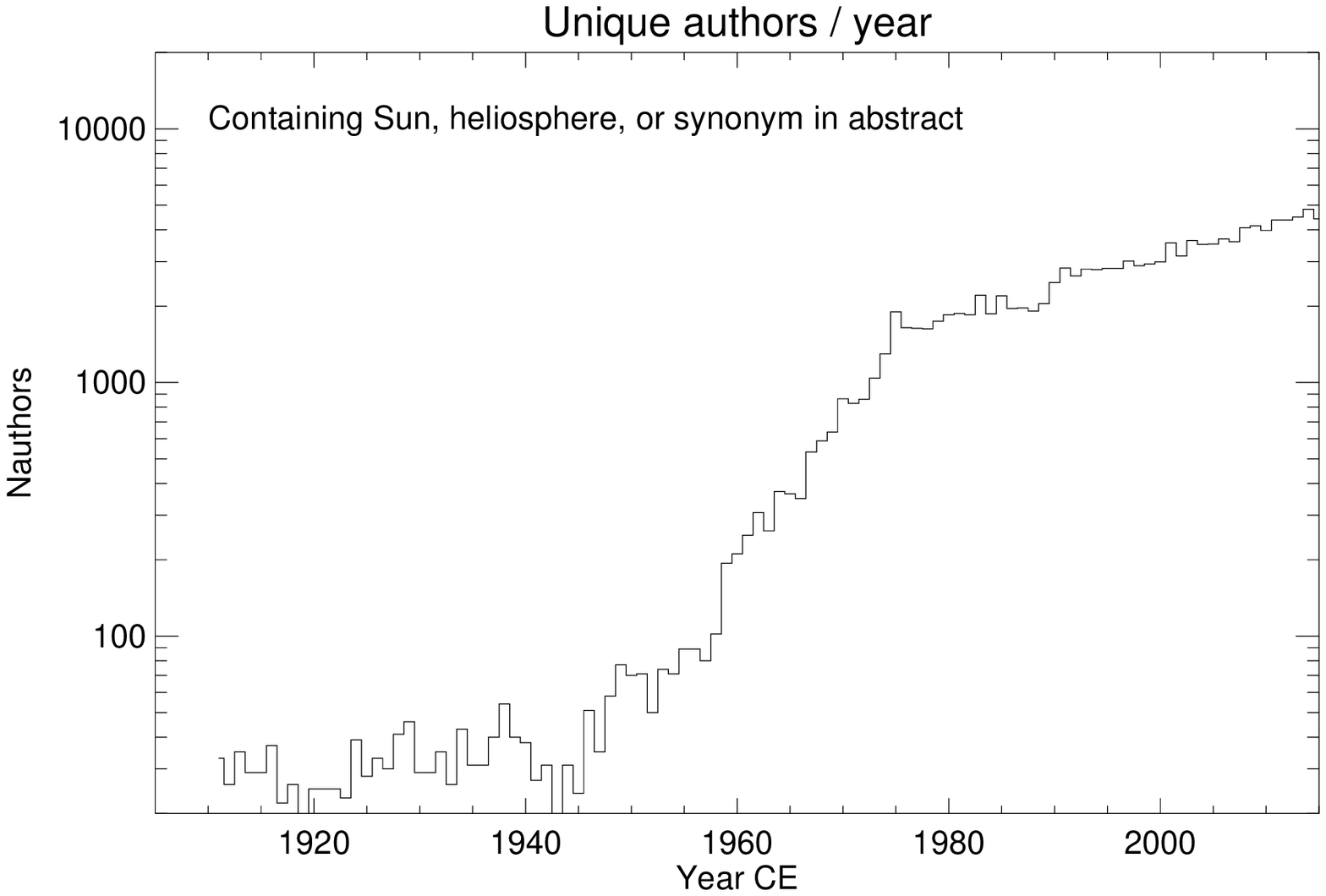}
\caption{Number of refereed publications per year with abstracts
  focusing on Sun or heliosphere {\em (top)} and number of unique
  authors for each year in publications per year with abstracts
  focusing on Sun or heliosphere {\em (bottom)}, as returned by
  ADS. (After Schrijver {\em et al.}, 2016, updated through 2015)}
\label{fig:nauthors}
\label{fig:yearlystats}
\end{figure}
The study of phenomena related to ``solar activity'' often involves
other aspects of the Sun (such as internal dynamics, dynamo, or
surface field patterns) and they are obviously not limited to the Sun
but drive phenomena throughout the heliosphere. \cite{comm10final} therefore do not
separate by research disciplines, but searched ADS for abstracts of refereed
publications in the ``Astronomy'' database, either mentioning the Sun
or heliosphere or their synonyms. They filtered out at least many of the
papers that do not deal with Sun/heliosphere that come into the search
results because their abstracts include, for example, a unit like
``solar mass''; to do so, they excluded abstracts that contain one or
more of the following words or word groups: cluster, dwarf,
extrasolar, galaxy, gravitational, ice, kpc, solar system, stellar,
binary, sunset, sunrise, eclipse, solar cell, solar occultation,
interstellar medium, and supernova.  

\begin{figure}[t]
\centering
\includegraphics[width=\textwidth]{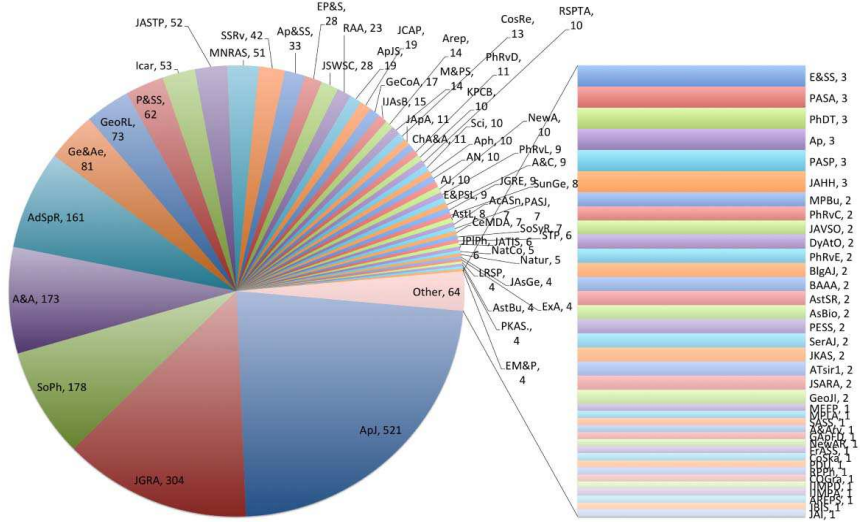}
\caption{Number of refereed publications in 2015 on solar or
  heliospheric topics sorted by refereed journal. The standard
  abbreviation of the journal name (\url{http://adsabs.harvard.edu/abs_doc/refereed1.html}) is followed by the number of
  selected publications \referee{(see caption to Fig.~3 for the names of the top 10)}.}
\label{fig:njournals}
\end{figure}
The ADS searches suggest that the productivity of the world-wide
community researching the Sun and heliosphere continues to grow
steadily when measured through its publications (top panel in
Figure~\ref{fig:yearlystats}). A rapid growth in the number of
refereed publications that started after the Second World War
continued up to about 1975. After that, the growth slowed drastically,
transitioning to a sustained increase that doubles the number of
refereed publications on a time scale of approximately 40 years,
reaching a total of 2285 publications by 2015.  \referee{This change
  (very similar to, for example, results obtained when looking for
  refereed publications with ``star'' or synomyms in the abstract; not
  shown here) is not likely to be an artefact of incompleteness of the
  ADS data bases, because these go back to first volumes of the most
  important journals, i.e., Astophysical Journal (from 1895 onward), Solar Physics
  (from 1967 onward), and Astronomy and Astrophysics (from 1969
  onward)\footnote{\url{http://adsabs.harvard.edu/journals_service.html}}.}

\begin{figure}[t]
\centering
\includegraphics[width=0.9\textwidth]{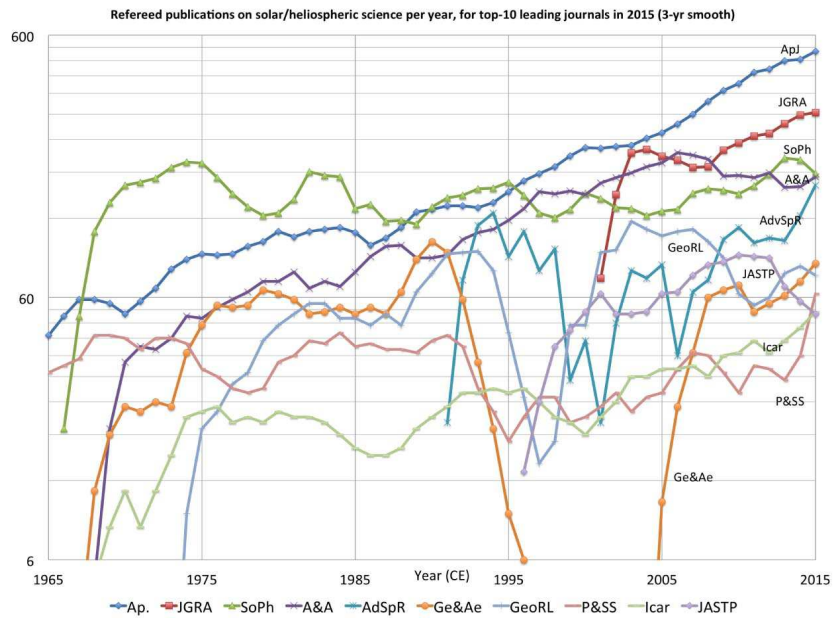}
\caption{Number of refereed publications per year (after 3-yr
  smoothing) on solar or heliospheric topics that appeared in the past
  50 years in the top-10 journals with most such publications in 2015:
  Astrophysical Journal (\referee{ApJ;} with 521 publications in
  2015), Journal of Geophysical Research Space Physics
  (\referee{JGRA;} 304 publications), Solar Physcis (\referee{SPh;}
  178), Astronomy and Astrophysics (\referee{A{\&}A;} 173), Advances
  in Space Research (\referee{AdSpR;} 161), Geomagnetism and Aeronomy
  (\referee{Ge{\&}Ae;} 81), Geophysical Research Letters
    (\referee{GeoRL;} 73), Planetary and Space Science
    (\referee{P{\&}SS;} 62), Icarus (\referee{Icar;} 53), and Journal
    of Atmospheric and Solar-Terrestrial Physics (\referee{JASTP;}
    52).}
\label{fig:journals}
\end{figure}
\referee{There are doubtlessly multiple reasons for the pronounced
  change in the publciation growth rate around 1975, but assessing
  these is beyond the scope of this paper. One substantial determining
  factor noted here, however, can be inferred from the U.S.\ federal
  budgets. Among other statistics, federal spending in nondefense
  research and development from 1953 onward as tracked by the
  AAAS\footnote{\url{http://www.aaas.org/page/historical-rd-data}}
  shows the rapid increase in post-WW\,II spending, peaking in the
  second half of the 1960s (including support for the fast-track
  Apollo project and its human exploration of the
  Moon), then decreasing to level off at a roughly constant funding
  when expressed in inflation-corrected units.}

\begin{table}
\caption{Number of refereed publications on solar
  and heliospheric physics grouped by country of the affiliation of the
  lead author, sorted by decreasing number. }\label{tab:countries}
\begin{tabular}{lr|lr|lr|lr}
\hline
USA & 567 & China & 262 & Russia & 228 &UK   & 126 \\
Germany & 112 & India & 110 &France & 84 & Japan & 74\\
Italy & 67 & Spain & 49 & Finland & 35 & Korea & 34\\
Belgium & 33 & Czech~Republ. & 28 & Norway & 27 & Sweden & 27\\
Brazil &24 & Ukraine & 23&Poland & 23 &Switzerland & 18\\
Austria & 16& Canada &15 & Australia &15&Greece &12\\
Mexico &11 &Bulgaria &10& South~Africa& 9&Nigeria &9\\
Argentina&8&Hungary&8&Iran&8&Thailand&8\\
Slovakia&7&Turkey&7&Taiwan&6&Ethiopia&6)\\
Egypt&5&Malaysia&4&Chile&4&Ireland&4\\
Denmark&4&Georgia&4&Israel&3&New~Zealand&3\\
the~Netherl.&3&Uganda&3&Croatia&3&Azarbaijan&2\\
Pakistan&2&Armenia&2&Romania&2&Portugal&2\\
Kenya&1&Algeria&1&Oman&1&Iraq&1\\
Slovenia&1&Venezuela&1&Cyprus&1&Saudi~Arabia&1\\
Costa~Rica&1&Colombia&1&Indonesia&1&Serbia&1\\
\hline
\end{tabular}
\end{table}
 In automated searches as used here it is not readily possible to avoid some
distortion of the statistics associated with the author names. For one
thing, authors with identical family names and initials for their
given names are not differentiated. Also, authors who publish with
different spellings or composites of their family names ({\it e.g.}
married and maiden names) or their initials will be counted as
separate individuals. However, these effects are expected to have
limited impacts on relative trends.

The number of unique author names contributing to refereed
publications shows a trend (shown in the bottom panel in
Figure~\ref{fig:yearlystats}) that roughly mimics that of the number of
publications; after about 1975, the growth rate of the author
population of $\approx 2.5$\%/yr is about twice the growth rate of the
world's overall population (which averaged at $\approx 1.3$\%/yr over
the same period; \citealp{prb2013}).

\section{Distribution over journals}
In 2015, ADS returned a total of 90 distinct refereed journals (plus a
few PhD theses that also qualify as refereed publications but are
likely underrepresented in the data base) with publications on solar
and heliospheric topics. Figure~\ref{fig:njournals} shows the
distribution over the journals, sorted by decreasing number of
publications in that year, using the standard abbreviations as used by
ADS\footnote{\url{http://adsabs.harvard.edu/abs_doc/refereed.html}}.

The top ten by number of publications in 2015 together contain about
73\%\ of all selected publications.  Figure~\ref{fig:journals} shows
the evolution of the number of papers in this top-10 over the past
half century (after a 3-yr smoothing to dampen fluctuations and for
easier viewing of the results). Note the relatively flat number of
publications in Solar Physics since about 1970 compared to the
sustained growth in that number in the Astrophysical Journal. Also
note that the predominant journal for solar and heliospheric science
changed from Solar Physics to the Astrophysical Journal around the
early 1990s, and that JGRA and A{\&}A overtook Solar Physics in the
first years of this millenium. Astronomy and Astrophysics shows a
relative growth in the rate of solar and heliospheric publications
similar to that in the Astrophysical Journal from the early 1970s
through 2006, after which a declining trend sets in over the past
decade.

\section{Nationalities of affiliations of lead authors}
ADS also enables a review of the affiliations of the lead authors of
the refereed publications in ADS for 2015 on solar and heliospheric
sciences and their use in other publications.  
When grouped by country for the affiliation of the lead author,
\referee{and sorted by decreasing
number of papers, the list of 64 countries results shown in 
Table~\ref{tab:countries}.} Funding agencies and national
Academies of science are involved through first authors in 
7.4\%\ of all papers from NASA in the USA, and
generally more than 50\%\ in China, Russia, and the former republics
of the USSR (where the Russian Academy of Sciences includes about
60\%\ of all scientific organizations).

\begin{figure}
\centering
\includegraphics[width=\textwidth]{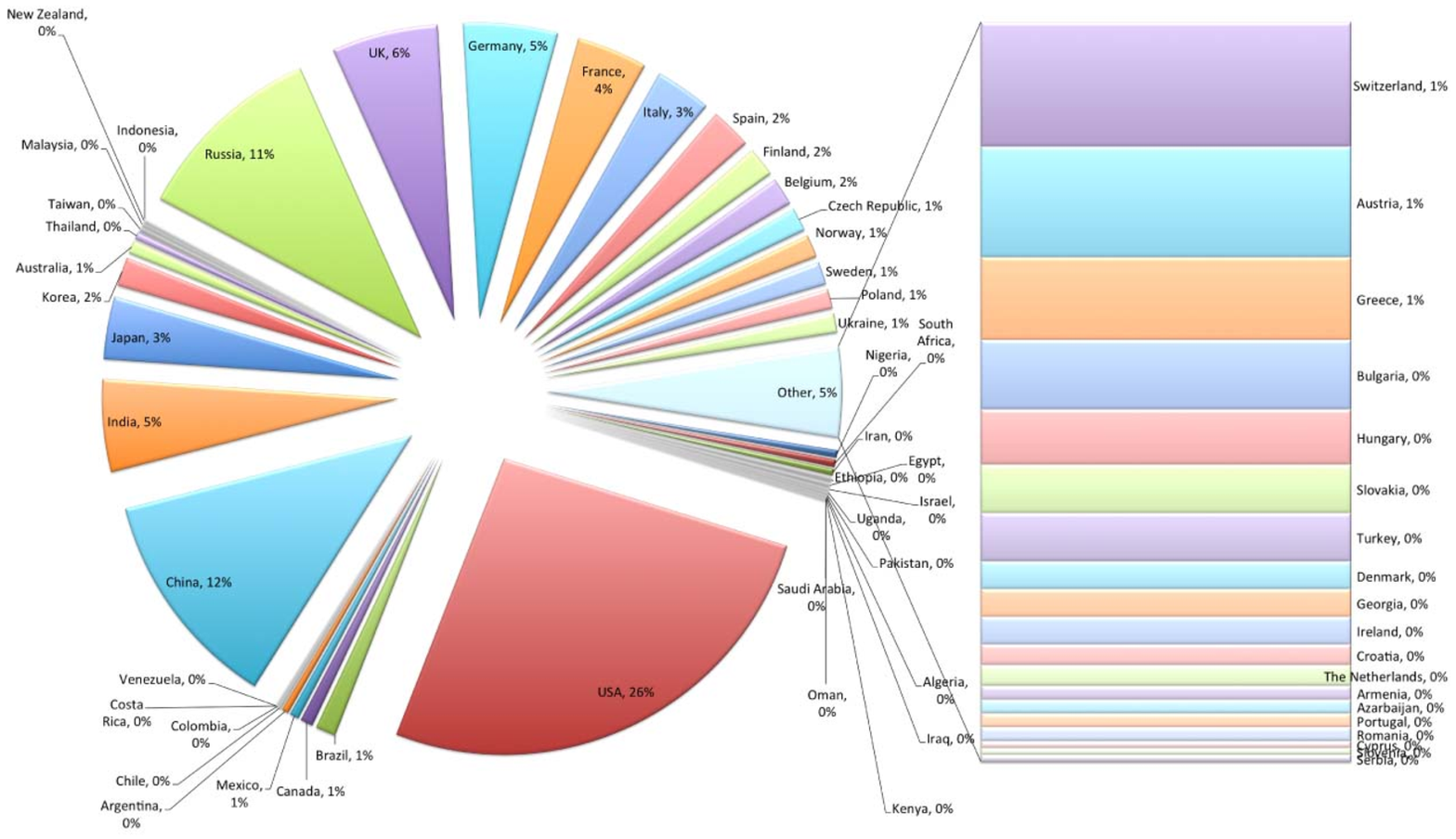}
\caption{Distribution of the number of refereed publications on solar
  and heliospheric physics grouped according to the country for the
  affiliation of the lead author, sorted by fraction of the total
  (shown in percent). The countries are shown sorted by general
  geographic region (clockwise from the top): Europe, Africa and the
  Arab region, Americas, and Australasia.}
\label{fig:regions}
\end{figure}
When sorting affiliations of lead authors by general geographic region
the following publications statistics result (see Figure~\ref{fig:regions}):
from Europe (including Russia): 969 (45\%\ of the total); from the
Americas: 632 (29\%); from Australasia: 517 (24\%); from Africa and
Arab countries (to Pakistan): 50 (2\%).




\end{article}
\end{document}